\def\papertitle{DiffVox: A Differentiable Model\\for Capturing and Analysing Vocal Effects Distributions}

\def\paperauthorA{Chin-Yun Yu}
\def\paperauthorB{Ben Hayes}
\def\paperauthorC{Gy\"orgy Fazekas}
\def\paperauthorD{Marco A. Martínez-Ramírez}
\def\paperauthorE{Junghyun Koo}
\def\paperauthorF{Wei-Hsiang Liao}
\def\paperauthorG{Yuki Mitsufuji}

\documentclass[twoside,a4paper]{article}
\usepackage{microtype}  
\usepackage{etoolbox}
\usepackage{dafx_25}
\usepackage{amsmath,amssymb,amsfonts,amsthm}
\usepackage{euscript}
\usepackage{mathtools}
\usepackage{booktabs}
\usepackage[T1]{fontenc}
\usepackage[utf8]{inputenc}
\usepackage{ifpdf}
\usepackage[english]{babel}
\usepackage{caption}
\usepackage{subfig} 
\usepackage{color, colortbl}
\usepackage{siunitx}
\usepackage{multirow}
\usepackage{enumitem}
\usepackage[htt]{hyphenat}
\usepackage[dvipsnames]{xcolor}

\input glyphtounicode
\ninept

\newcounter{numauth}
\newcounter{listcnt}
\newcommand\authcnt[1]{\ifdefined#1 \stepcounter{numauth} \fi}

\newcommand\addauth[1]{
\ifdefined#1 
\stepcounter{listcnt}
\ifnum \value{listcnt}<\value{numauth}
\appto\authorslist{, #1}
\else
\appto\authorslist{~and~#1}
\fi
\fi}
\authcnt{\paperauthorB}
\authcnt{\paperauthorC}
\authcnt{\paperauthorD}
\authcnt{\paperauthorE}
\authcnt{\paperauthorF}
\authcnt{\paperauthorG}
\authcnt{\paperauthorH}
\authcnt{\paperauthorI}
\authcnt{\paperauthorJ}
\def\authorslist{\paperauthorA}
\addauth{\paperauthorB}
\addauth{\paperauthorC}
\addauth{\paperauthorD}
\addauth{\paperauthorE}
\addauth{\paperauthorF}
\addauth{\paperauthorG}
\addauth{\paperauthorH}
\addauth{\paperauthorI}
\addauth{\paperauthorJ}

\usepackage{times}

\newif\ifpdf
\ifx\pdfoutput\relax
\else
   \ifcase\pdfoutput
      \pdffalse
   \else
      \pdftrue
   \fi
\fi

\ifpdf 
  \usepackage[pdftex,
    pdftitle={\papertitle},
    pdfauthor={\authorslist},
    pdfsubject={Proceedings of the 28th International Conference on Digital Audio Effects (DAFx25)},
    colorlinks=false, 
    bookmarksnumbered, 
    pdfstartview=XYZ 
  ]{hyperref}
  \usepackage[pdftex]{graphicx}
\else 
  \usepackage[dvips]{epsfig,graphicx}
  \usepackage[dvips,
    pdftitle={\papertitle},
    pdfauthor={\authorslist},
    pdfsubject={Proceedings of the 28th International Conference on Digital Audio Effects (DAFx25)},
    colorlinks=false, 
    bookmarksnumbered, 
    pdfstartview=XYZ 
  ]{hyperref}
\fi
\usepackage[hypcap=true]{caption}
\title{\papertitle}

\affiliation{
\paperauthorA$^\flat$\sthanks{Work done during an internship at Sony AI.}, 
\paperauthorD$^\sharp$, \paperauthorE$^\sharp$, \paperauthorB$^\flat$, 
\paperauthorF$^\sharp$, \paperauthorC$^\flat$, and \paperauthorG$^{\sharp\natural}$}
{$^\flat$\href{https://www.c4dm.eecs.qmul.ac.uk/}{Centre for Digital Music}, Queen Mary University of London, London, UK\\
$^\sharp$\href{https://ai.sony/}{Sony AI}, Tokyo, Japan\\
$^\natural$\href{https://www.sony.net/}{Sony Group Corporation}, Tokyo, Japan\\
{\tt \href{mailto:chin-yun.yu@qmul.ac.uk}{chin-yun.yu@qmul.ac.uk}}
}

\begin{document}
\ifpdf 
  \DeclareGraphicsExtensions{.png,.jpg,.pdf}
\else  
  \DeclareGraphicsExtensions{.eps}
\fi

\setlength{\abovedisplayskip}{5pt}
\setlength{\belowdisplayskip}{5pt}
\setlength{\parfillskip}{0pt plus 0.5\columnwidth}

\maketitle

\begin{abstract}
  This study introduces a novel and interpretable model, DiffVox, for matching vocal effects in music production.
  DiffVox, short for ``\textbf{Diff}erentiable \textbf{Vo}cal F\textbf{x}", integrates parametric equalisation, dynamic range control, delay, and reverb with efficient differentiable implementations to enable gradient-based optimisation for parameter estimation.
  Vocal presets are retrieved from two datasets, comprising 70 tracks from MedleyDB and 365 tracks from a private collection.
  Analysis of parameter correlations reveals strong relationships between effects and parameters, such as the high-pass and low-shelf filters often working together to shape the low end, and the delay time correlating with the intensity of the delayed signals.
  Principal component analysis reveals connections to McAdams' timbre dimensions, where the most crucial component modulates the perceived spaciousness while the secondary components influence spectral brightness.
  Statistical testing confirms the non-Gaussian nature of the parameter distribution, highlighting the complexity of the vocal effects space.
  These initial findings on the parameter distributions set the foundation for future research in vocal effects modelling and automatic mixing.
\end{abstract}

\section{Introduction}
\label{sec:intro}

Audio effects are essential in music production.
They enable audio engineers to shape a sound's timbre and spatial characteristics, such as stereo width.
Understanding how these effects (their settings) are used in real-world audio is valuable for developing automatic audio processing tools to create realistic music mixes.
Yet, this knowledge is based on decades of experience and often remains untracked systematically.
Since the distribution of effect parameters is unknown, we often approximate it with non-informative priors, such as uniform or Gaussian distributions, to ease downstream tasks.
This is often used for generating synthetic training data for a classifier model that identifies the applied processing, including tasks such as effects detection~\cite{rice2023general}, music mixing style transfer systems~\cite{steinmetz2022style,tony_style_2023,koo2025itomasterinferencetimeoptimizationaudio}, or pretraining audio representations~\cite{christian_j_steinmetz_2024_14877423}.
This forms a biased and weighted training objective, where the effect settings that are less likely to occur in real-world mixes are overrepresented, and vice versa, thus cancelling out the prior.
The influence of weighting has been found in filter design using neural networks~\cite{colonel2022direct}, where different sampling strategies of the filter coefficients affect the generalisation results on real-world impulse responses (IRs).

Unfortunately, due to the complexity of the music production process, it is challenging to collect real-world data with annotations on the effects parameters.
Reverse-engineering the mix is more feasible.
Directly optimising the effects parameters end-to-end using gradient descent is effective in equaliser matching~\cite{nercessian2020neural}, fitting IRs with filters~\cite{bhattacharya2020optimization} or feedback delay networks (FDN)~\cite{mezza2024data}, learning compressor parameters~\cite{wright2022grey}, or even capturing the whole mixing process graph~\cite{lee2025reverse}.
We thus adopt differentiable sound matching as a proxy to capture real-world effects configurations.
For simplicity and feasibility, we focus on a \textbf{single track, mono-in-stereo-out} scenario, independent of broader mix interactions between tracks.
We choose vocals because they are often the most prominent element in a mix and are carefully processed.
The resulting parameters can be considered as \emph{presets} sampled from arbitrary mixes.
The order and routing of the effects are fixed, resulting in fixed dimensionality, which makes later analysis tractable.
We follow~\cite{lee2025reverse} in using static parameters to control the effects, which is sufficient to represent complete mixes.

Our contributions are as follows:
Firstly, we propose an effects model that reflects professional music production practices while being efficient to train in a differentiable manner.
We incorporate parallel algorithms running on graphics processing units (GPUs) to accelerate fitting the recursive filters used in the equaliser and dynamic range controller.
We implement a differentiable ping-pong delay, an FDN reverb with frequency-dependent attenuation, and a dynamic range controller with look-ahead.
Secondly, we propose a loss function that matches the signals' microdynamics in a multi-resolution fashion, similar to common spectral losses, to capture the features of interest in different scales.
Thirdly, we fit the effects to hundreds of vocal tracks and analyse the collected presets.
Our analysis reveals the importance of spatial effects for sound matching, highlights strong correlations of specific parameters, and demonstrates that the most explainable components of the parameter distributions are related to spaciousness and spectral brightness.
Lastly, we publicise our experiments' source code and the vocal presets dataset to foster further research on audio effects prior~\footnote{\href{https://github.com/SonyResearch/diffvox}{github.com/SonyResearch/diffvox}}.

\begin{figure*}[ht!]
  \centering
  \includegraphics[width=0.98\textwidth]{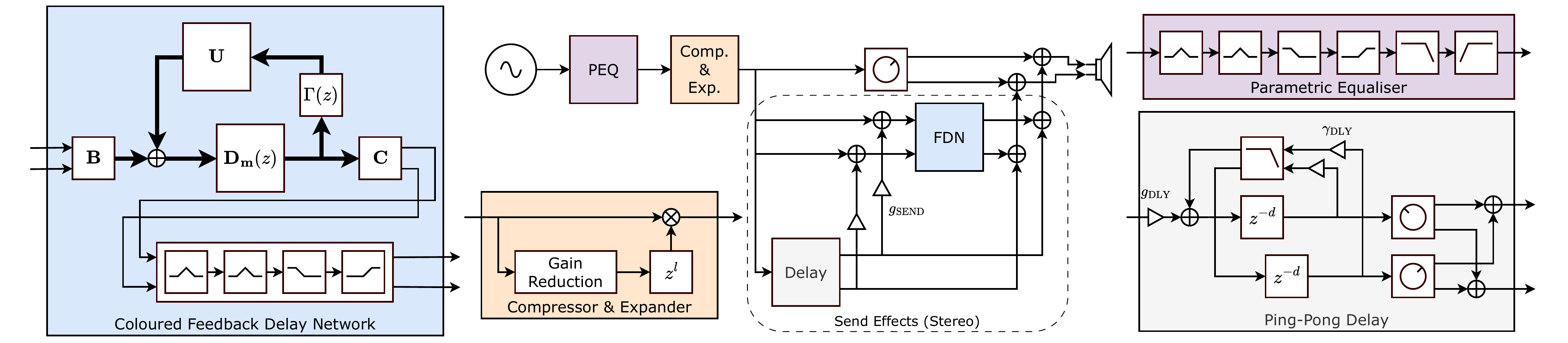}
  \caption{The proposed model (upper left) and individual effects for vocal effects processing.}
  \label{fig:diagram}
\end{figure*}

\section{The Effects Model}
\label{sec:model}
Our chosen effects are based on standard practices in music production~\footnote{\href{https://www.soundonsound.com/techniques/vocal-production}{www.soundonsound.com/techniques/vocal-production}}.
The mono input is first treated with a six-band parametric equaliser, followed by a compressor and an expander as a dynamic range controller.
Then, the signal is split into two paths: one for the dry signal and the other for the wet signal.
A ping-pong delay and an FDN reverb process the wet signal.
A panner processes the dry signal and then mixes it with the wet signal.
The exact routes are shown in Fig.~\ref{fig:diagram}.
We pick effect implementations with the fewest parameters possible to reduce the dimensionality so it does not exceed the number of vocal tracks we collected.
We approximate the effects only when necessary to reduce fitting time.
In the following sections, we describe each effect in detail.

\subsection{The Parametric Equaliser (PEQ)}
\label{ssec:eq}

The PEQ sequentially applies the following six filters: two peak filters (PK1 and PK2), a low-shelf filter (LS), a high-shelf filter (HS), a low-pass filter (LP), and a high-pass filter (HP).
We reference the T-RackS Classic Equaliser plugin by IK Multimedia\footnote{\href{https://www.ikmultimedia.com/products/trclasseq/}{www.ikmultimedia.com/products/trclasseq}} to set the ranges of the parameters.
We follow the Audio EQ Cookbook\footnote{\href{https://www.w3.org/TR/audio-eq-cookbook/}{https://www.w3.org/TR/audio-eq-cookbook/}} to implement the filters as Biquad filters since they are commonly used in digital audio effects.
We fix the shelf filters' Q factor to 0.707, resulting in four gains, four Q factors, and six frequencies to be optimised.

Recently, differentiable time-domain evaluation of time-\linebreak invariant recursive filters has been made possible by specialised kernels enabling efficient backpropagation~\cite{ycy2024golfv1}.
However, these kernels do not employ any parallelisation along the time axis, meaning they do not fully utilise parallel processors.
Blelloch's parallel prefix sum algorithm~\cite{BlellochTR90} illustrates that recursive expressions can be parallelised if the recursion can be expressed in terms of an associative operation.
This reduces time complexity from $O(N)$ to $\approx O(\frac{N}{p})$, where $p$ is the number of parallel processors.
Thus, we seek to express the Biquad filter recursion using an associative operation, allowing us to apply the parallel scan algorithm~\cite{harris2007parallel}.

{\sloppy
A Biquad filter consists of five parameters $\{b_0, b_1, b_2, a_1, a_2\}$.
The state-space realisation~\cite{FILTERS07} of a Biquad filter in Direct-Form-II is:
\begin{equation}
  \label{eq:ssr}
  \begin{aligned}
    \tilde{\bf x}[n+1] & = \mathbf{A}_{\rm BQ}\tilde{\bf x}[n] + \begin{bmatrix}
                                                                   x[n] \\ 0
                                                                 \end{bmatrix} \\
    y[n]               & = \mathbf{C}_{\rm BQ}\tilde{\bf x}[n] + b_0 x[n],
  \end{aligned}
\end{equation}
where $\mathbf{A}_{\rm BQ} = \begin{bmatrix} -a_1 & -a_2 \\ 1 & 0 \end{bmatrix}$ and $\mathbf{C}_{\rm BQ} = \big[b_1 - b_0 a_1 \quad b_2 - b_0 a_2\big]$.
Noting that the first line of Eq.~\eqref{eq:ssr} can be rewritten without recursion:
\begin{equation}
  \tilde{\bf x}[n] = \sum_{k=1}^{n} \mathbf{A}_{\rm BQ}^{n-k}
  \begin{bmatrix}
    x[k - 1] \\ 0
  \end{bmatrix},
\end{equation}
\noindent we define an associative binary operation $\oplus$ acting on tuples $(\mathbf{U}_{1}, \mathbf{v}_{1}) \oplus (\mathbf{U}_{2}, \mathbf{v}_{2}) \mapsto (\mathbf{U}_{2}\mathbf{U}_{1}, \mathbf{U}_{2}\mathbf{v}_{1} + \mathbf{v}_{2})$.
By setting $\mathbf{U}_{n} = \mathbf{A}_{\rm BQ}$, $\mathbf{v}_n = \begin{bmatrix}
    x[n - 1] & 0
  \end{bmatrix}^\top$, and $s_n = (\mathbf{U}_{n}, \mathbf{v}_{n})$, we can express the recursion step of our filter associatively:
\begin{equation}
  \begin{aligned}
    \tilde{s}_n & = s_1 \oplus s_2 \oplus \dots \oplus s_n = \bigoplus_{k=1}^n \left(
    \mathbf{A}_{\rm BQ},
    \begin{bmatrix} x[k - 1] \\ 0 \end{bmatrix}
    \right)
  \end{aligned},
\end{equation}
where $\tilde{s}_n^{(2)}$, the second entry of the tuple $\tilde{s}_n$, gives us $\tilde{\bf x}[n]$.
Moreover, if $\mathbf{A}_{\rm BQ}$ is diagonalisable, applying $\oplus$ can be simplified further, reducing matrix multiplications to scalar multiplications.
If the poles of the filter $\lambda_1, \lambda_2$ are distinct, $\mathbf{A}_{\rm BQ}$ can be diagonalised as
$\mathbf{P} \mathbf{\Lambda}\mathbf{P}^{-1}$ and $\mathbf{\Lambda} = \text{diag}(\begin{bmatrix} \lambda_1 & \lambda_2 \end{bmatrix})$.
Using the fact that $\mathbf{A}_{\rm BQ}^n = \mathbf{P} \mathbf{\Lambda}^n\mathbf{P}^{-1}$ and altering our associative representation such that $\mathbf{v}_n = \mathbf{P}^{-1}\begin{bmatrix}
    x[n - 1] & 0
  \end{bmatrix}^\top$ and, accordingly,
$\mathbf{U}_n = \mathbf{\Lambda}$,
we recover our filtered signal $\tilde{\mathbf{x}}[n] = \mathbf{P}\tilde{s}_n^{(2)}$.}

To ensure distinct poles, we restrict the Q factor of the HP and LP filters to be no smaller than 0.5.
In the case of real poles, we set $\mathbf{P} = \begin{bmatrix} \begin{bmatrix}1 & {\lambda_1}^{-1} \end{bmatrix}^\top & \begin{bmatrix} 1& {\lambda_2}^{-1} \end{bmatrix}^\top \end{bmatrix}$.
For complex conjugate poles ($\lambda_1 = \lambda_2^*$), we utilise the coupled form state-space model~\cite{laroche_stability_2007}.
In the coupled form, the transition matrix is a rotation matrix.
Applying the rotation matrix to a two-dimensional vector is equivalent to complex multiplication where the multiplier is the pole $\lambda_1$.
In other words, we can run just one complex one-pole filter instead of two.
The rotation and Direct-Form-II transition matrices are interchangeable by the following equations:
\begin{equation}
  \label{eq:coupled}
  \mathbf{A}_{\rm BQ}  = \mathbf{P} \begin{bmatrix}\Re(\lambda_1) & -\Im(\lambda_1) \\ \Im(\lambda_1) & \Re(\lambda_1) \end{bmatrix} \mathbf{P}^{-1}, \,
  \mathbf{P}^{-1} = \begin{bmatrix} 0 & \Im(\lambda_1) \\ -1 & \Re(\lambda_1) \end{bmatrix}.
\end{equation}
Plug Eq.~\eqref{eq:coupled} into the first line of Eq.~\eqref{eq:ssr}, multiply $\mathbf{P}^{-1}$ on both sides, and convert every vector into a complex number, we get the following recursion:
\begin{equation}
  \label{eq:ssr_cplx}
  \begin{aligned}
    \bar{x}[n+1] & = \lambda_1 \bar{x}[n] - i x[n]. \\
  \end{aligned}
\end{equation}
Let $\mathbf{v}_n = -i x[n-1]$ and $\mathbf{U}_n = \lambda_1$ in the associative operation, then the filtered signal $\tilde{\mathbf{x}}[n]$ is $\mathbf{P}\begin{bmatrix} \Re(\tilde{s}_n^{(2)}) & \Im(\tilde{s}_n^{(2)}) \end{bmatrix}^\top$.
The computational cost is the same as in Eq.~\eqref{eq:ssr}, but the implementation is more straightforward since we can treat it as a one-pole filter.

\subsection{The Feed-Forward Compressor and Expander (COMP)}
\label{ssec:comp}
We adopt the compressor and expander model from the DAFx textbook~\cite{dafx_comp}.
The effect is controlled by the following parameters: the compressor/expander thresholds $CT/ET$, the compressor/expander ratios $CR/ER$, the attack/release/RMS smoothing factors $\alpha_{\rm at}/\alpha_{\rm rt}/\alpha_{\rm rms}$, and the make-up gain.
We use the differentiable implementation \texttt{torchcomp} by Yu et al.~\cite{ycy2024diffapf}.
The RMS level detector and the backpropagation of the attack/release ballistic filter are also one-pole filters; thus, we accelerate them using the parallel scan.

Due to the causal smoothing effect of the RMS level detector, the gain reduction signal $g[n]$ is slightly delayed.
Modern digital compressors often have a look-ahead feature to compensate for this delay.
To learn the continuous delay time $l \in \mathbb{R}_+$, we approximate the look-ahead by truncated sinc interpolation:
\begin{equation}
  \label{eq:interp}
  g(n+l) \approx \sum_{k=-L_1}^{L_2+1} g[n+k]\text{sinc}\left(k - l\right),
\end{equation}
where $l \in [0, L_2]$ and $L_1, L_2$ are the truncation lengths.
The sinc function is defined as $x \mapsto \frac{\sin(\pi x)}{\pi x}$.
The output of the compressor is then $y_{\rm COMP}[n] = g(n+l) x[n]$.

\subsection{The Ping-Pong Delay (DLY: $\mathbb{R} \to \mathbb{R}^2$)}
\label{ssec:pingpong}
Ping-Pong delay is a stereo delay effect where the delays alternate between the left and right channels.
It is implemented by two delay lines whose outputs are each other's inputs.
In modern music production, the panning of the delay is more flexible and not limited to hard left and right.
We thus add two separate panners to control the panning of two delay lines.
We add an LP filter in the feedback path to simulate the decaying echoes.
The LP filter is the same Biquad filter as in Section~\ref{ssec:eq}.

We adopt the damped sinusoidal approach to learn the delay time using frequency-sampling (FS)~\cite{hayes2023sinusoidal,lee_grafx_2024}, representing the delay effects as convolutions with a truncated finite impulse response (FIR) $\mathbf{h}_{\rm DLY}[n]$ with a length of $N_{\rm DLY}$.
The following equations approximate the transfer functions of the delays:
\begin{align}
   & \begin{multlined}
       H_{\rm odd}(z)   = \frac{z^{-d}}{1 - \gamma_{\rm DLY} H_{\rm LP}(z)z^{-2d}}                                                                                                                                     \\
       \approx \gamma_{\rm DLY}^{-1}\sum_{k=1}^{\left\lfloor{\frac{N_{\rm DLY}-d}{2d}}\right\rfloor}\left(\gamma_{\rm DLY}H_{\rm LP}(z)\eta^{\frac{\angle{z}}{2\pi}N_{\rm DLY}}z^{-d}\right)^{2k-1},
       \label{eq:pingpong_odd}
     \end{multlined} \\
   & \begin{multlined}
       H_{\rm even}(z) = \frac{\gamma_{\rm DLY} H_{\rm LP}(z)z^{-2d}}{1 - \gamma_{\rm DLY} H_{\rm LP}(z)z^{-2d}} \\
       \approx \sum_{k=1}^{\left\lfloor{\frac{N_{\rm DLY}}{2d}}\right\rfloor - 1}\left(\gamma_{\rm DLY}H_{\rm LP}(z)\eta^{\frac{\angle{z}}{2\pi}N_{\rm DLY}}z^{-d}\right)^{2k},
       \label{eq:pingpong_even}
     \end{multlined}
\end{align}
where $\eta \in [0, 1]$ is the surrogate variable, $\gamma_{\rm DLY} \in [0, 1]$ is the decay factor, $d$ is the delay time, and $H_{\rm LP}(z)$ is the transfer function of the LP filter.
$\eta^{\frac{\angle{z}}{2\pi}N_{\rm DLY}}z^{-d}$ forms a damped sinusoidal in the frequency domain, a surrogate way to learn the true delay operator $z^{-d}$.
We set $\eta = 1$ during inference, assuming the variable always converges to one.
We explicitly compute each delayed impulse within the range $0 \leq n < N_{\rm DLY}$ to reduce the time aliasing effect when FS the original transfer function~\cite{lee_reverb_2022}.

The two impulses are then joined together as
\begin{equation}
  \label{eq:pingpong_join}
  \mathbf{h}_{\rm DLY}[n] = \text{PAN}_{\rm odd}(h_{\rm odd}[n]) + \text{PAN}_{\rm even}(h_{\rm even}[n])
\end{equation}
where $\text{PAN}: \mathbb{R} \to \mathbb{R}^2$ is the panning function.
Following~\cite{lee_grafx_2024}, we use the straight-through estimator to backpropagate the gradients to the damped sinusoidal.
The final output of the ping-pong delay is the convolution of the input signal with the IR $\mathbf{y}_{\rm DLY}[n] = g_{\rm DLY}(y_{\rm COMP}[n] \ast \mathbf{h}_{\rm DLY}[n])$
and $g_{\rm DLY} \in [0, 1]$ controls the volume of the delayed signal.

\subsection{The Feedback Delay Network Reverb (FDN: $\mathbb{R}^2 \to \mathbb{R}^2$)}
\label{ssec:fdn}
FDN is an artificial reverberation algorithm that uses a network of delay lines with feedback to create dense reverberations~\cite{stautner1982designing}.
In this work, we use a stereo FDN with six delay lines.
It is best described in state-space form as:
\begin{equation}
  \label{eq:fdn}
  \begin{aligned}
    \begin{bmatrix}
      \tilde{x}_1[n+m_1] \\
      \tilde{x}_2[n+m_2] \\
      \vdots             \\
      \tilde{x}_6[n+m_6] \\
    \end{bmatrix}   & = \mathbf{A}\tilde{\mathbf{x}}[n] + \mathbf{B}\mathbf{x}[n], \\
    \mathbf{y}_{\rm FDN}[n] & = \mathbf{C}\tilde{\mathbf{x}}[n],
  \end{aligned}
\end{equation}
where $\mathbf{A} \in \mathbb{R}^{6 \times 6}$, $\mathbf{B} \in \mathbb{R}^{6 \times 2}$, $\mathbf{C} \in \mathbb{R}^{2 \times 6}$, and $m_i$ are the delay times.
$\mathbf{A}$ controls how the energies spread across delay lines.
The delay times are co-primes to increase echo density.
We set the delay times to $\mathbf{m}$ = [997, 1153, 1327, 1559, 1801, 2099] proposed in~\cite{dal2023differentiable}.

Computing the recursion of Eq.~\eqref{eq:fdn} directly in automatic differentiation frameworks is time-consuming due to the overhead of an enormous amount of function calls to register computational nodes, and each contributes little computation~\cite{ycy2024golfv1}.
Furthermore, there are no specialised kernels for Eq.~\eqref{eq:fdn} (in contrast to the PEQ in Section~\ref{ssec:eq}); thus, we use the FS method to approximate the IR of the FDN.
The transfer function $\mathbf{H}_{\rm FDN}(z) = \frac{\mathbf{Y}_{\rm FDN}(z)}{\mathbf{X}(z)}$ is given by the following equation:
\begin{gather}
  \label{eq:fdn_tf}
  \mathbf{H}_{\rm FDN}(z) = \mathbf{C}\left(\mathbf{D}_\mathbf{m}^{-1}(z) - \mathbf{A}(z)\right)^{-1}\mathbf{B},\\
  \mathbf{D}_\mathbf{m}(z) = \text{diag}\left(\begin{bmatrix} z^{-m_1} & z^{-m_2} & \ldots & z^{-m_6} \end{bmatrix}\right).
\end{gather}
We parametrise $\mathbf{A}(z) = \mathbf{U}\Gamma(z)$ where $\mathbf{U}$ is an orthogonal matrix and $\Gamma(z) = \text{diag}([\gamma^{m_1}(z) \ldots \gamma^{m_6}(z)]), \gamma(z) \in [0, 1]$.
We follow Mezza et al.~\cite{mezza2024data} to parametrise $\mathbf{U}$ to be unitary, thus satisfying the unilosslessness condition~\cite{schlecht2016lossless}.
$\gamma(z)$ is the attenuation filter that controls the decay rate of the reverb.

In most of the previous works on differentiable FDN~\cite{mezza2024data, dal2023differentiable, giampiccolo2024differentiable}, frequency-independent $\gamma$ is used to parametrise $\mathbf{A}$, which limits the flexibility of the model.
The difficulty is that we usually want the decay time to be delay-independent, and designing such filters is non-trivial.
Mezza et al.~\cite{mezza2024modeling} tackle this problem indirectly by learning separate FIRs for each delay line and training them with a frequency-dependent objective.
Here, we adopt a more straightforward approach that aligns with the FS method.
We sample 49 points~\footnote{The number was decided empirically based on early experiments by reducing the number of points until it loses spectrum details.} of $\gamma(z)$ with equal spacing from 0 to $\pi$.
The attenuation coefficients are then upsampled to the desired length of the FFT during FS.
After fitting, we can approximate the FDN by calculating the linear-phase filter from the magnitude response of $\gamma^{m_i}(z)$ and then applying Mezza's model for real-time purposes.
In addition, the decaying time of the reverb in Eq.~\eqref{eq:fdn_tf} is frequency-dependent, while the initial gain of the reverb is not.
To correct this, we add a PEQ after the reverb, which contains two peak filters and two shelf filters from Section~\ref{ssec:eq}.
In practice, we apply PEQ on the impulse response of the reverb $\mathbf{h}_{\rm FDN}[n]$ before convolving it with the input signal for efficiency.

\subsection{The Effect Sends (SEND) and Parametrisation}
\label{ssec:sends}
Both the reverb (Section \ref{ssec:fdn}) and the delay (Section \ref{ssec:pingpong}) only model the wet signal.
To further increase the model's flexibility, we also send the delayed signal to the reverb to colourise the delays to have similar acoustic characteristics to the direct signal, controlled by the send level $g_{\rm SEND} \in [0, 1]$.
The total number of parameters in our effect chain is 152\footnote{14 for PEQ, 9 for the compressor and expander, 8 for the delay, 119 for the FDN reverb, one for the panning, and one for the send level.}.
Compared to GRAFx~\cite{lee2025reverse, lee_grafx_2024}, a package that provides differentiable effects modelling, a similar effects signal chain requires at least 264 more parameters~\footnote{This number is based on replacing our delay and reverbs with theirs.}, since in GRAFx the goal is to represent the mix faithfully, thus their delay and reverbs are over-parametrised to be expressive enough for various mixing materials.
In contrast, our model specialises in vocals, and we prioritise having a compact representation of parameters for analysis over expressiveness.
Since many effects' parameter ranges are bounded, we apply different parametrisation to the parameters, which are summarised in Table~\ref{tab:param}.

\begin{table*}[ht!]
  \caption{The parametrisation of the effects. $\text{tri}(\mathbf{X})$ is the upper triangular part of the matrix $\mathbf{X}$. For details on the bounds, please refer to our code repository.}
  \centering
    \begin{tabular}{ccc}
    \toprule
    Condition ($\mathbb{P}$)                 & Parametrisation ($\mathbb{R} \to \mathbb{P}$)                                                            & Parameters ($\theta \in \mathbb{R}$)                                                                           \\
    \midrule
    $x \in \mathbb{R}$                       & $\theta \mapsto \theta$                                                                                  & Equaliser's/make-up gain, $CT, ET, \mathbf{B}, \mathbf{C}$                                                     \\
    $0 \leq x \leq 1$                        & $\sigma: \theta \mapsto \frac{1}{1 + e^{-\theta}}$                                                       & Panning, $\alpha_{\rm rms}, \alpha_{\rm at}, \alpha_{\rm rt}, ER, \gamma_{\rm DLY}, g_{\rm DLY}, g_{\rm SEND}$ \\
    $a \leq x \leq b$                        & $\theta \mapsto a + \sigma(\theta) (b - a)$                                                              & Equaliser's Q and frequency, $CR, d, \gamma(z)$                                                                \\
    $0 \leq x \leq a$                        & $\theta \mapsto |\theta| \mod a$                                                                         & $l$                                                                                                            \\
        $x \leq 0$                               & $\theta \mapsto -\log (1 + e^\theta)$                                                                    & $\log(\eta)$                                                                                                   \\
    $\mathbf{X}^\top\mathbf{X} = \mathbf{I}$ & $\boldsymbol{\Theta} \mapsto e^{\text{tri}(\boldsymbol{\Theta}) - \text{tri}(\boldsymbol{\Theta})^\top}$ & $\mathbf{U}$                                                                                                   \\
    \bottomrule
  \end{tabular}
  \label{tab:param}
\end{table*}

\section{Experiments}
\label{sec:exp}

\subsection{Datasets}
\label{ssec:data}
We apply our effects model to two datasets: 1) the \texttt{MedleyDB}~\cite{bittner2014medleydb,bittner2016medleydb} and 2) our private multi-track dataset \texttt{Internal} with paired dry and wet stems~\cite{lee2025reverse}.
The latter consists mainly of modern mainstream Western music.
Both datasets are sampled at \SI{44.1}{\kHz}.
We use the official metadata of \texttt{MedleyDB} to pick the vocal tracks.
For \texttt{Internal}, the pairing information between raw tracks and processed stems is missing.
We calculate the cross-correlations between each song's dry tracks and wet stems and use these correlations to recover their mapping.
We then drop non-vocal stems based on their filenames.

Stems processed from only one raw track are selected to fit our problem setting (\emph{mono-in-stereo-out}).
Some input tracks are stereo, possibly due to the exporting process from the DAW.
For these tracks, we first peak-normalise both channels and then calculate their difference (side channel).
We drop the track if the maximum side energy exceeds \SI{-10}{\decibel}.
We then take the average of the two channels to form a mono source.
Since the raw tracks and processed stems are not always aligned in time, we time-align the raw tracks so their cross-correlation with the processed stems is maximised.

\subsection{Optimisation}
\label{ssec:opt}

The loss functions we use are 1) the multi-scale STFT (MSS) loss, 2) the multi-scale Loudness Dynamic Range (MLDR) loss, and 3) the regularisation loss on the surrogate variable $\eta$.
The MSS loss is defined as
\begin{multline}
  \label{eq:ms}
  \mathcal{L}_{\text{MSS}}(\hat{y}[n], y[n])
  = \frac{1}{3}\sum_{N \in \{128, 512, 2048\}} \frac{\|\hat{\mathbf{Y}}_N - \mathbf{Y}_N\|_2}{\|\mathbf{Y}_N\|_2} \\
  + \frac{\|\log(\hat{\mathbf{Y}}_N) - \log(\mathbf{Y}_N)\|_1}{NM_N},
\end{multline}
where $\hat{\mathbf{Y}}_N$ and $\mathbf{Y}_N$ are the magnitude spectrograms of the predicted and ground-truth signals, respectively, computed with FFT size $N$ and hop size $\frac{N}{4}$.
$M_N$ is the number of frames in the spectrogram.
Similar to~\cite{lee2025reverse}, we use \texttt{auraloss}\footnote{\href{https://github.com/csteinmetz1/auraloss}{github.com/csteinmetz1/auraloss}} to compute the MSS loss, with A-weighting applied before the STFT~\cite{wright2020perceptual}.
This loss minimises the distance in the spectral domain.

Inspired by previous work on differentiable microdynamics metrics~\cite{nercessian_direct_2022}, we propose the MLDR loss to match the dynamics of the predicted and ground-truth signals, thus better guiding the fitting of the compressor.
Given a signal $x[n]$, its LDR is defined as
\begin{multline}
  \label{eq:ldr}
  \text{LDR}(x[n], t_{\rm short}, t_{\rm long}) \\
  = \log\left(\frac{\text{RMS}\left(x^2[n], t_{\rm short}\right)}{\text{RMS}\left(x^2[n + \lfloor \frac{t_{\rm long} - t_{\rm short}}{2T_s}\rfloor], t_{\rm long}\right)}\right),
\end{multline}
where RMS calculates the energy envelopes of the signal, $t_{\rm short}$ and $t_{\rm long}$ are the integration times in second, and $T_s$ is the sampling period.
The longer the integration time, the smoother the RMS envelope.
For more calculation details, please refer to~\cite{nercessian_direct_2022}.
$t_{\rm long} \gg t_{\rm short}$, so the LDR describes how the loudness varies locally in a scale proportional to $t_{\rm long}^{-1}$.
Similar to MSS loss, we want to match the LDR from coarse to fine scales.
Thus, we propose the following $L_1$ loss:
\begin{multline}
  \label{eq:mldr}
  \mathcal{L}_{\text{MLDR}}(\hat{y}[n], y[n]) \\
  = \frac{1}{N}\sum_{t \in \{1, 2\}} \sum_{n=0}^{N-1}\left|\text{LDR}\left(\hat{y}[n], \frac{t}{20}, t\right) - \text{LDR}\left(y[n], \frac{t}{20}, t\right)\right|
\end{multline}
where $N$ is the length of the signal.

The final loss function is the weighted sum of the MSS and MLDR losses on the left, right, mid, and side channels, plus the regularisation loss on $\eta$:
\begin{multline}
  \label{eq:loss}
  \mathcal{L}(\hat{\mathbf{y}}[n], \mathbf{y}[n]) = \frac{1}{2} \sum_{i=1}^2 \Big[ \Big.\mathcal{L}_{\text{MSS}}(\hat{y}_i[n], y_i[n]) \\
  + \frac{1}{2}\mathcal{L}_{\text{MSS}}(\sqrt{2}\mathcal{H}(\hat{\mathbf{y}}[n])_i, \sqrt{2}\mathcal{H}(\mathbf{y}[n])_i)
  + \frac{1}{2}\mathcal{L}_{\text{MLDR}}(\hat{y}_i[n], y_i[n]) \\
  + \frac{1}{4}  \mathcal{L}_{\text{MLDR}}(\mathcal{H}(\hat{\mathbf{y}}[n])_i, \mathcal{H}(\mathbf{y}[n])_i)\Big. \Big] + (1 - \eta)^2,
\end{multline}
where $\mathcal{H}: \begin{bmatrix}
    x & y
  \end{bmatrix}^\top \mapsto \frac{1}{\sqrt{2}}\begin{bmatrix}
    x + y & x - y
  \end{bmatrix}^\top$ is the Hadamard transform that converts left/right to mid/side channels.
The last term encourages the damped sinusoidal to be on the unit circle.
The weights are set empirically to balance the initial magnitude of the individual losses.

\subsection{Fitting Details}
\label{ssec:train}

We normalise the input and target vocals to have \SI{-18}{\decibel} LUFS~\cite{lufs} using \texttt{pyloudnorm}\footnote{\href{https://github.com/csteinmetz1/pyloudnorm}{github.com/csteinmetz1/pyloudnorm}}.
Each track is then split into twelve-second segments with five seconds of overlap.
The overlapped region is used as a warm-up, and the loss is calculated only for the last seven seconds, similar to~\cite{lee2025reverse}.
We drop silent segments and select up to 35 segments in each training step to form a batch.
We train the effects on each track for 2k steps using the Adam optimiser with 0.01 learning rate and pick the checkpoint with the lowest loss.
We use the CUDA implementation of parallel scan by Martin et al.~\cite{martin2018parallelizing}.
The fitting time of each track ranges from 20 to 40 minutes on a single RTX 3090 GPU.

The PK and LS/HS filters in the PEQ are initialised with zero gains.
The PK filters' cut-off frequencies are bounded differently, so their parameters are ordered and not permutation-invariant.
The cut-off frequencies for LP and HP filters are initialised to \SI{17.5}{\kHz} and \SI{200}{\hertz}, respectively.
The dynamic range controls are initialised with $CR=2, ER=\frac{1}{2}, CT=-18, ET=-48$, and make-up gain \SI{0}{\decibel}.
We initialise the delay with $\gamma_{\rm DLY}=g_{\rm DLY}=0.1$ and delay time to \SI{400}{\ms}.
We initialise the FDN with $\mathbf{B} = \mathbf{1}, \mathbf{C} = \mathbf{0}, \gamma(z) \sim \mathcal{U}(0.4, 0.6)$.
The send level is initialised to 0.01.
The model is initialised very close to an identity function, which is the upper bound for the loss, so we can easily detect problematic runs.
We set the impulse response length to four seconds for the delay and 12 seconds for the FDN reverb.
We bound the damping factor $\gamma(z)$ to have a maximum of nine seconds T60 time to reduce the aliasing effect, but still be long enough to capture most of the reverb tail.

A few tracks have non-linear effects, such as distortion and modulations, that are not modelled in our effects.
To exclude them, we drop fitting runs that 1) have a minimum loss above a certain threshold, 2) have a loss that fluctuates heavily during fitting, or 3) have a loss that is not decreasing.
The thresholds are set empirically after checking runs with apparent outlier fitting losses.
Out of 76 tracks from \texttt{MedleyDB}, 6 tracks are excluded ($\approx 8\%$); out of 370 tracks from \texttt{Internal}, 5 tracks are excluded ($\approx 1.3\%$).

\section{Results and Discussions}
\label{sec:res}

\subsection{Sound Matching Performance}
\label{ssec:match}
We fit different configurations to test the benefits of having spatial effects.
We denote the complete model as DiffVox.
The evaluation metrics are the fitting losses on the left/right and mid/side channels.
The averaged scores across tracks are shown in Table~\ref{tab:comparison}.

\begin{table}[h]
  \caption{Matching performance with different configurations.}
  \centering
  \resizebox{0.98\columnwidth}{!}{
    \begin{tabular}{llcccc}
      \toprule
      \bf \multirow{2}{*}{Dataset}       & \multirow{2}{*}{\bf Configuration}  & \multicolumn{2}{c}{\bf MSS $\downarrow$} & \multicolumn{2}{c}{\bf MLDR $\downarrow$}                       \\
                                         &                                     & l/r                                      & m/s                                       & l/r      & m/s      \\
      \midrule
      \multirow{3}{*}{\texttt{Internal}} & No processing                       & 1.44                                     & 2.39                                      & 1.82     & 2.08     \\
      \cmidrule{2-6}
                                         & DiffVox                             & \bf 0.76                                 & \bf 0.94                                  & \bf 0.34 & \bf 0.41 \\
                                         & ~\rotatebox{90}{$\top$} w/o Approx. & 0.78                                     & 0.95                                      & 0.38     & 0.44     \\
      \midrule
      \multirow{7}{*}{\texttt{MedleyDB}} & No processing                       & 1.27                                     & 2.16                                      & 1.00     & 1.35
      \\
      \cmidrule{2-6}
                                         & DiffVox                             & 0.75                                     & 0.98                                      & \bf 0.39 & \bf 0.45 \\
                                         & ~\rotatebox{90}{$\top$} w/o Approx. & 0.77                                     & 1.00                                      & 0.42     & 0.48     \\
                                               & w/o FDN                             & 0.79                                     & 1.14                                      & 0.48     & 0.62     \\
                                         & w/o DLY                             & 0.76                                     & 0.99                                      & 0.40     & 0.47     \\
                                         & w/o DLY\&FDN                        & \bf 0.61                                 & \bf 0.90                                  & 0.82     & 1.17     \\
      
      \bottomrule
    \end{tabular}
  }
  \label{tab:comparison}
\end{table}

From Table~\ref{tab:comparison}, we can see that with just PEQ and compressor, the rendered audio matches well regarding spectral content, as indicated by the MSS on \texttt{MedleyDB}.
However, it fails to match the microdynamics indicated by the MLDR loss.
Adding the delay or the FDN improves the matching of the microdynamics.
DiffVox achieves the best matching performance in MLDR and a lower MSS than with just Delay or FDN, proving its ability to match the sound in spectral and dynamic aspects.
After removing the approximation, the performance drops slightly, which is expected since we swap truncated FIRs of reverb and delay with infinite IRs during inference, allowing the longer tails in the IRs to introduce a mismatch.
Nevertheless, it is still better than without FDN.

\subsection{Parameter Correlation Analysis}
\label{ssec:corr}

We analyse the correlation between the parameters in both datasets.
Specifically, on the minimum set of parameters to reproduce the effects $\boldsymbol{\theta} \in \mathbb{R}^{130}$, which excludes the surrogate variable $\eta$ and the lower triangular part of the logits of $\mathbf{U}$.
Since the parameters are unlikely to be normally distributed, we compute their Spearman correlation coefficient (SCC) instead of regular correlations\footnote{For convenience, the analysis is performed on the parameter logits before the parametrisation. This does not affect SCC for most of the scalar parameters (excludes $\mathbf{U}$ and $l$) since their parametrisation in Table~\ref{tab:param} is monotonic.}.
We exclude the correlations of multi-dimensional parameters $\gamma(z)$, $\mathbf{B}$, $\mathbf{C}$, $\mathbf{U}$ since interpreting them is not straightforward.
We observe high correlations between the sampled $\gamma(z)$, suggesting that the attenuation coefficients can be decomposed into fewer parameters.

The following discussions are based on the top ten most correlated pairs of parameters.
\texttt{MedleyDB}'s most correlated pairs are not aligned with \texttt{Internal}, showing that the two datasets have different characteristics.
Since \texttt{MedleyDB} has less data compared to \texttt{Internal} (70 vs. 365 tracks), the correlation we see is also less reliable.
Thus, we primarily focus on the \texttt{Internal} dataset.

High correlations are observed between the delay effect parameters.
The negative correlations between the delay time and the feedback gain (-.58) and the delay gain (-.51) imply that when a longer delay time is used, the delay effects diminish.
The positive correlation between the feedback gain and the cut-off frequency of the delay's LP filter (.49) indicates that when a darker delay is used, it also fades out faster, and vice versa.
In addition, high correlations are observed between the gain and the Q factor of the PEQ's PK2 filter (-.60) and the gain and cut-off frequency of the FDN's PK2 filter (-.46).
The counter-interaction between the compressor threshold and the make-up gain is also observed (-.55), which is expected, as the two parameters are usually adjusted together to ensure consistency in loudness before and after the compressor.

We see that the attenuation coefficients above \SI{19.7}{\kHz} are highly correlated (ranging from .53 to .60) with the LP filter cut-off frequency.
The reverb tends to compensate for the high-frequency loss by reducing the decay rate because we set the maximum cut-off frequency to \SI{18}{\kHz} according to the T-RackS EQ.
A possible solution is to set a higher bound for the cut-off frequency or incorporate a wet/dry mix ratio on LP similar to GRAFx~\cite{lee_grafx_2024}.

To analyse the correlation from a broader perspective, we measure the correlations between individual effects in the effects model on \texttt{Internal}.
We define the effect-wise correlation as the average of the absolute SCCs between the parameters of the two effects.
For correlations within the same effect, the diagonal elements are excluded.
We see that most of the effects have high autocorrelation, except for the PEQ's LS filter (.011).
The LS filter has a high correlation with the HP filter (.198).
The LS filter only has two parameters: gain and frequency.
A low correlation means that the two parameters operate nearly independently.
This implies the two filters are used collaboratively to shape the low end.
Based on the correlations, the hierarchical clustering using Ward's method~\cite{ward1963hierarchical} reveals three main clusters: all the spatial effects, the HP and LS filters, and the remaining effects.
These clusters could be used to inform a simpler design of vocal effects.

\subsection{Principal Component Analysis}
\label{ssec:pca}
Inspired by a similar dataset work~\cite{guezenoc_wide_2020}, we perform principal component analysis (PCA)~\cite{pca2002} on the parameters' logits to analyse the distribution of the parameters.
Although PCA is limited to linear relationships, its simplicity and interpretability make it a good starting point for understanding the effects prior.
Given the sample covariance matrix $\mathbf{\Sigma}_{\boldsymbol{\theta}} \in \mathbb{R}^{130 \times 130}$ of the logits $\boldsymbol{\theta}$, we compute the eigenvalues $\boldsymbol{\lambda}_{\boldsymbol{\theta}} \in \mathbb{R}^{r}$ and the eigenvectors $\mathbf{V}_{\boldsymbol{\theta}} \in \mathbb{R}^{130 \times r}$ such that
\begin{equation}
  \begin{split}
    \mathbf{\Sigma}_{\boldsymbol{\theta}}
    = \mathbf{V}_{\boldsymbol{\theta}}
    \text{diag}(\begin{bmatrix}\lambda_{\boldsymbol{\theta}_1} & \lambda_{\boldsymbol{\theta}_2} & \ldots &\lambda_{\boldsymbol{\theta}_{r}}\end{bmatrix})
    \mathbf{V}_{\boldsymbol{\theta}}^\top, \\
    \lambda_{\boldsymbol{\theta}_1} \geq \lambda_{\boldsymbol{\theta}_2} \geq \ldots \geq \lambda_{\boldsymbol{\theta}_{r}} > 0.
  \end{split}
\end{equation}
$\mathbf{V}_{\boldsymbol{\theta}}$ is an orthogonal matrix, and the eigenvectors are the principal components (PCs) of the parameters.
The eigenvalues represent the variance of the parameters in the direction of the PCs.
$r$ is the rank of $\mathbf{\Sigma}_{\boldsymbol{\theta}}$ and equals the number of non-zero eigenvalues.

\begin{figure}[h]
  \centering
  \includegraphics[width=\columnwidth]{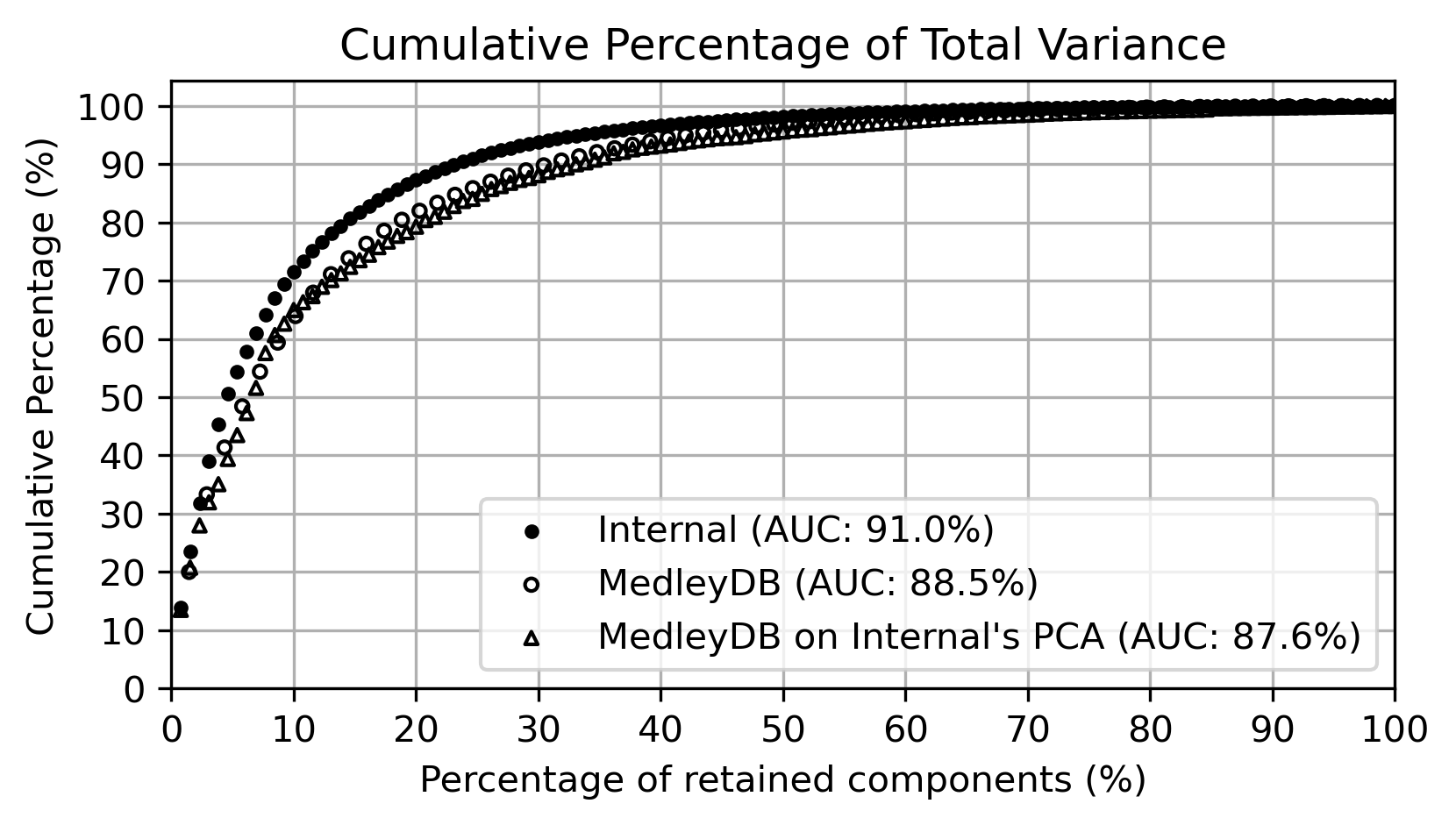}
  \caption{Cumulative total variance as a function of the percentage of retained PCs ($100p/r$) from both PCA models.}
  \label{fig:pca_cpv}
\end{figure}

The cumulative percentage of total variance (CPV, \cite{pca2002}) is a simple method for evaluating the PCA model's capacity.
The CPV given $p$ number of components to retain is defined as $100 \sum_{i=1}^p \lambda_{\boldsymbol{\theta}_i}$\linebreak$/\sum_{i=1}^{r} \lambda_{\boldsymbol{\theta}_i}$.
The CPV of both models is shown in Fig.~\ref{fig:pca_cpv}.
\texttt{Internal} has more variance explained by the first $p$ components than \texttt{MedleyDB}, indicated by the larger area under the curve (AUC = \SI{91.0}{\percent} vs \SI{88.5}{\percent}).
In other words, the parameters in \texttt{Internal} are more densely distributed than \texttt{MedleyDB}.
To see how \texttt{MedleyDB} is explained by the \texttt{Internal}'s PCA model, we also compute another CPV by replacing the eigenvalues with the sum of squared projections of \texttt{MedleyDB}'s parameters onto each \texttt{Internal}'s PC\footnote{The PCA projection is defined as $\mathbf{x} \mapsto \mathbf{V}_{\boldsymbol{\theta}}^\top (\mathbf{x} - \boldsymbol{\mu}_{\boldsymbol{\theta}})$ where $\mathbf{\mu}_{\boldsymbol{\theta}}$ is the sample mean of the parameters.} and plot it in Fig.~\ref{fig:pca_cpv}.
It shows that $\approx$ \SI{65}{\percent} of the variances in \texttt{MedleyDB} can be captured by \texttt{Internal}'s first ten PCs, and the CPV trends are similar, but for higher PCs, the discrepancy increases.

One application of having a PCA model is using it as a generative model, drawing samples from the distribution $\mathcal{N}(\textbf{0}, \text{diag}(\boldsymbol{\lambda}_{\boldsymbol{\theta}}))$.
To test this assumption, we perform two multivariate normality tests, Royston's~\cite{royston1983} and Henze-Zirkler's~\cite{Henze01011990} tests, on the PC weights of \texttt{Internal}.
The p-values of the tests are nearly zero given the first 75 PCs ($\approx$ \SI{99}{\percent} of the total variance), indicating that the logits are unlikely to be drawn from a multivariate normal distribution.

\begin{figure*}[!t]
  \centering
  \includegraphics[width=\textwidth]{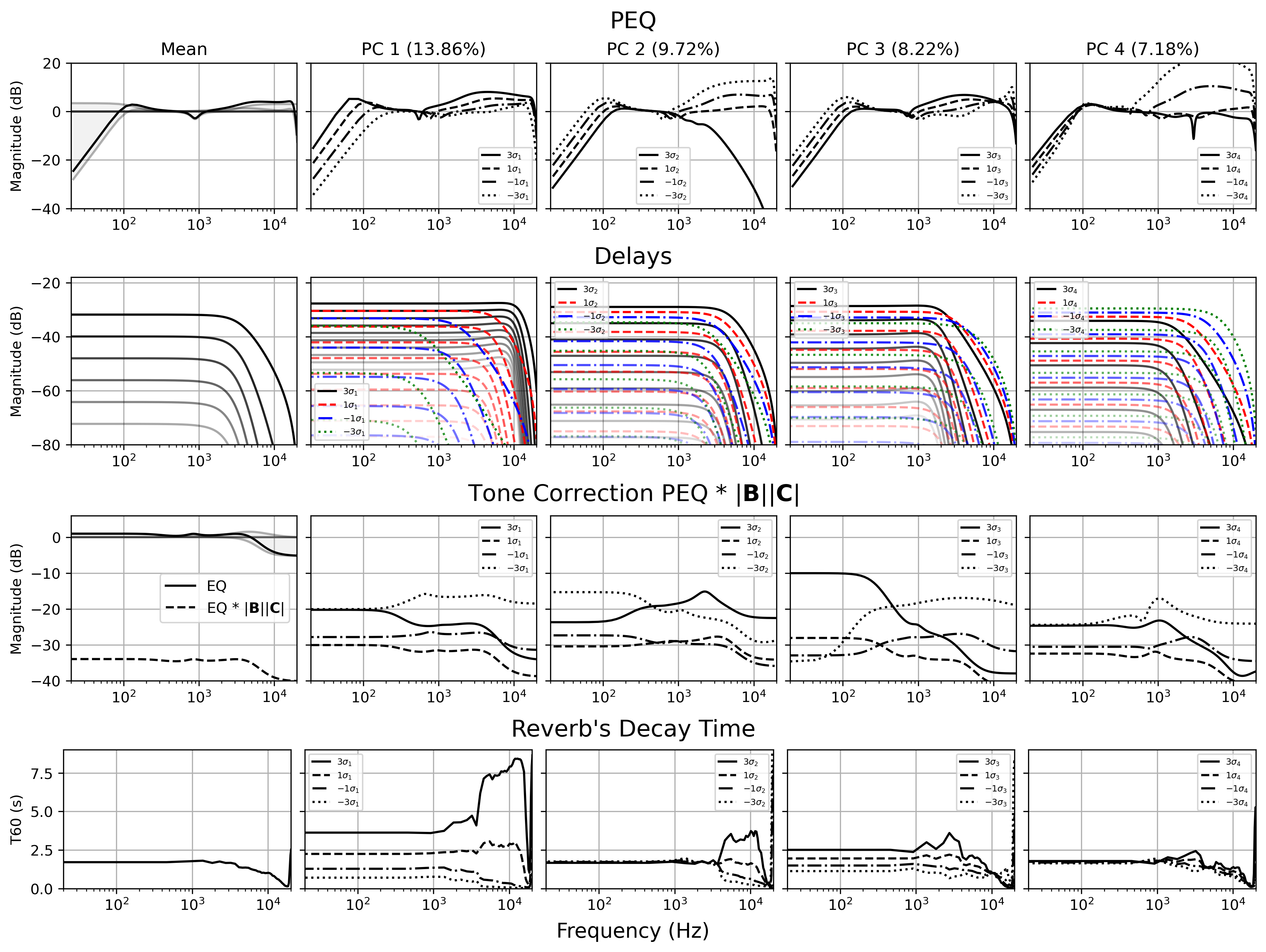}
  \caption{
    The mean (column one) and the first four principal components (column two to five with their percentage of explained variance) of the \texttt{Internal} dataset.
    The first and third rows show the frequency response of the PEQ and the tone correction filter. The second row shows the frequency response of the delayed signals, with colour intensity proportional to the delay time. The fourth row shows the frequency-dependent decay of the FDN reverb.
  }
  \label{fig:pc_compare}
\end{figure*}

We plot the frequency responses of the PEQ, the delay, the tone correction PEQ, and the reverb decay time in Fig.~\ref{fig:pc_compare}.
The mean parameters (first column) show a reasonable configuration that we could expect for a professionally processed vocal.
Both the high and low ends are boosted, and the mid-range close to \SI{1}{\kHz} is slightly attenuated with narrow Q.
An adequate amount of delay and reverb is added, with the high frequencies in the reverb attenuated.
The reverb's decay time is around 2 seconds, starting at low frequencies and gradually decreasing to zero.
The compression ratio is set to 3.5:1, with a threshold of \SI{-22}{\decibel} and a make-up gain of \SI{2.4}{\decibel}.
The vocals are polished and professionally processed.

To see how the PCs affect the parameters, we add the $i^{\rm th}$ PC to $\boldsymbol{\mu}_{\boldsymbol{\theta}}$ with scales in $\{3, 1, -1, -3\} \times \sqrt{\lambda}_{\boldsymbol{\theta}_i}$.
The first PC (second column) mainly affects the spaciousness of the acoustic property where the vocals are placed.
The feedback gain and overall volume of the delay are increased.
The decay time of the reverb is also significantly increased, especially in frequencies above \SI{4}{\kHz}, which creates a very shimmery and spacious sound.
The second PC (third column) creates a band-pass effect similar to that of a telephone, which can be seen from the drastic changes in the HS and LP filters.
This aligns with McAdams' timbre space~\cite{mcadams1995perceptual}, where their second dimension is related to the spectral centroid.
Long reverberations also smooth the attack time, which is related to McAdams' first dimension.
The dynamic range compression gets slightly heavier in the fourth PC but does not change much in the first three.

\section{Conclusions}
\label{sec:conclude}

This paper presents an expressive differentiable model for vocal effects processing and a method for capturing the effects parameters from professional mixes.
Spatial effects are crucial for achieving good matching performance, as indicated by lower microdynamics losses.
The parameter correlation analysis reveals meaningful relationships between effect parameters, such as the interaction between delay time and feedback gain and between compressor threshold and make-up gain.
The first two principal components of the PCA model on our private dataset reveal primary directions that alter the vocals, including control of spaciousness and frequency bandwidth manipulation (telephone-like effects), which have some connections to McAdams' timbre space.

While our approach successfully captures effect parameters for most tracks, limitations remain in handling vocals with other effects we did not model or heavy automation.
Addressing these limits and extending the model to multi-track scenarios is left for future work.
The non-normality of the parameter distribution suggests that a more sophisticated generative model is needed to capture the actual distribution.
We release the dataset of 435 vocal preset parameter logits produced in this study, plus a test set of 20 additional presets from the \texttt{Internal} dataset.
The dataset is accompanied by the DiffVox model implemented in PyTorch, scripts to reproduce the results on the \texttt{MedleyDB} dataset, and the PCA models.
We hope these resources can advance the development of future automatic mixing tools or neural audio effect models.

\section{Acknowledgments}
We thank Sungho Lee for providing the cross-correlation data of the \texttt{Internal} dataset.
This research is supported jointly by UKRI (grant number EP/S022694/1) and QMUL and utilised Queen Mary's Apocrita HPC facility, supported by QMUL Research-IT. \href{http://doi.org/10.5281/zenodo.438045}{doi: 10.5281/zenodo.438045}

\bibliographystyle{IEEEbib}
\bibliography{ref} 
\appendix

\section{Supplementary Results}
\label{sec:figures}

\begin{figure}[h]
    \centering
    \includegraphics[width=\columnwidth]{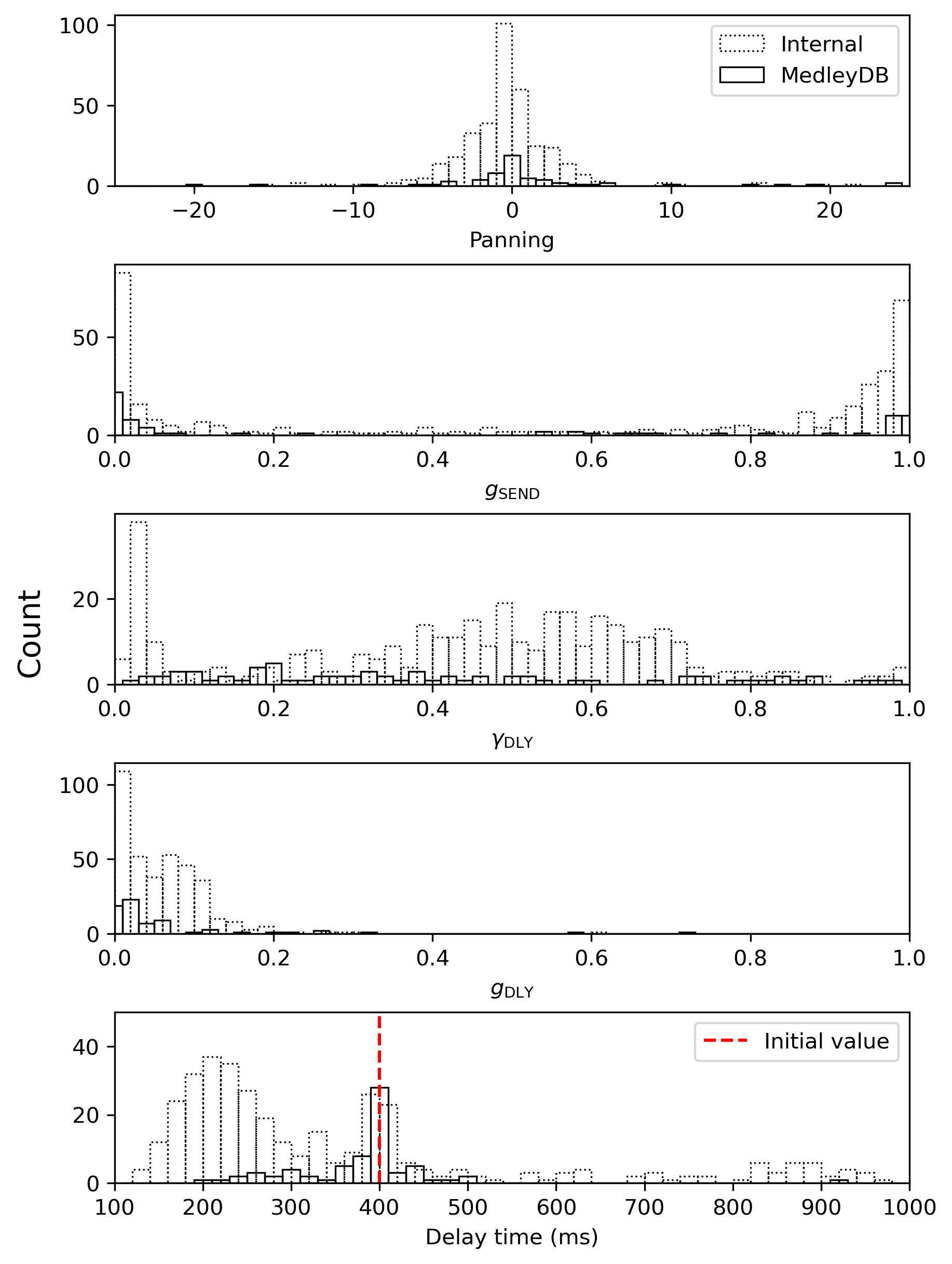}
    \caption{The histograms of the panning, delay send, and delay parameters.}
    \label{fig:panning_send}
\end{figure}

\begin{figure}[h]
    \centering
    \includegraphics[width=\columnwidth]{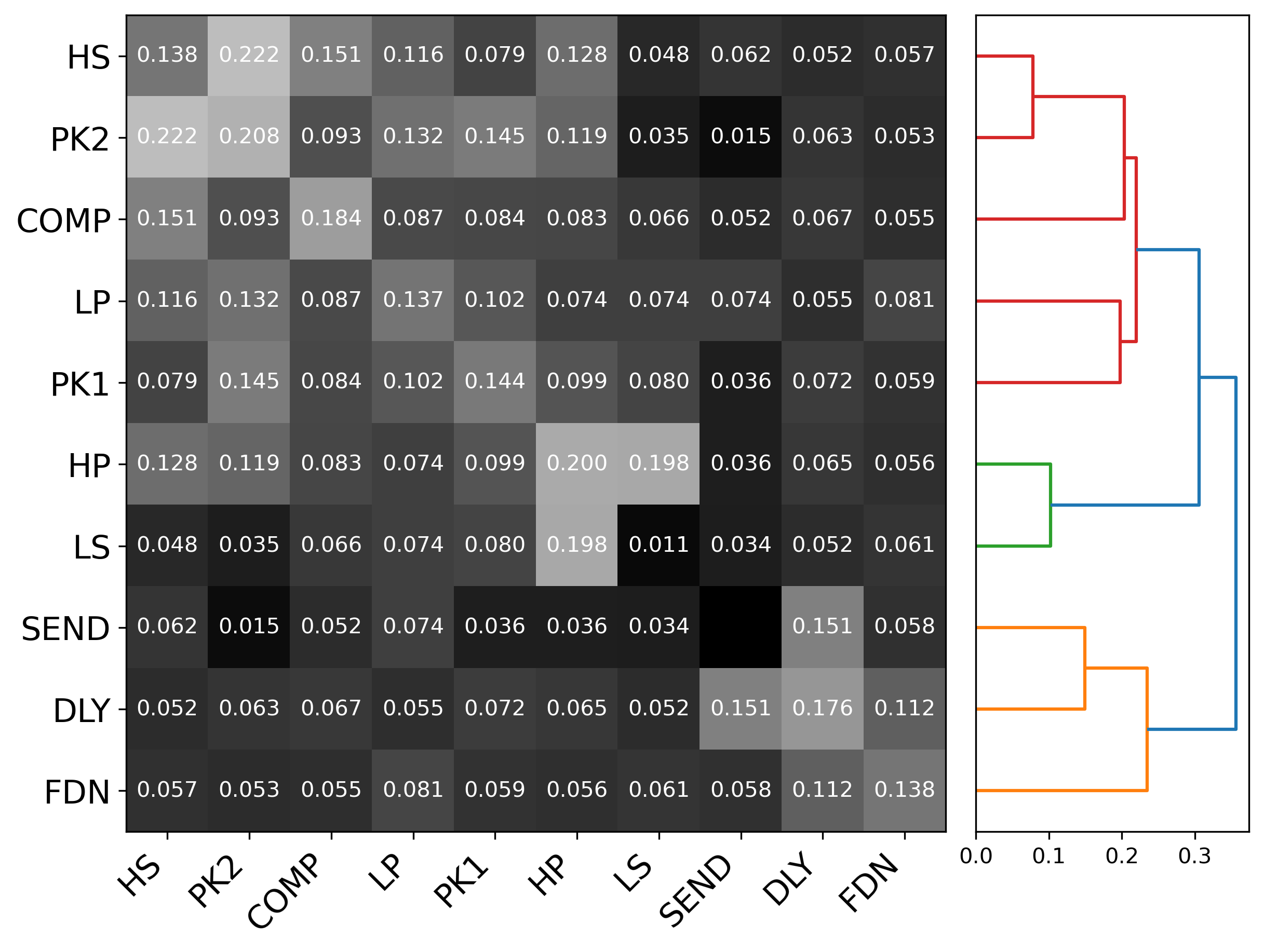}
    \caption{Left: The correlation matrix of \texttt{Internal}'s effects. Right: The hierarchical clustering.}
    \label{fig:effect_corr}
\end{figure}

\begin{figure}[h]
    \centering
    \includegraphics[width=\columnwidth]{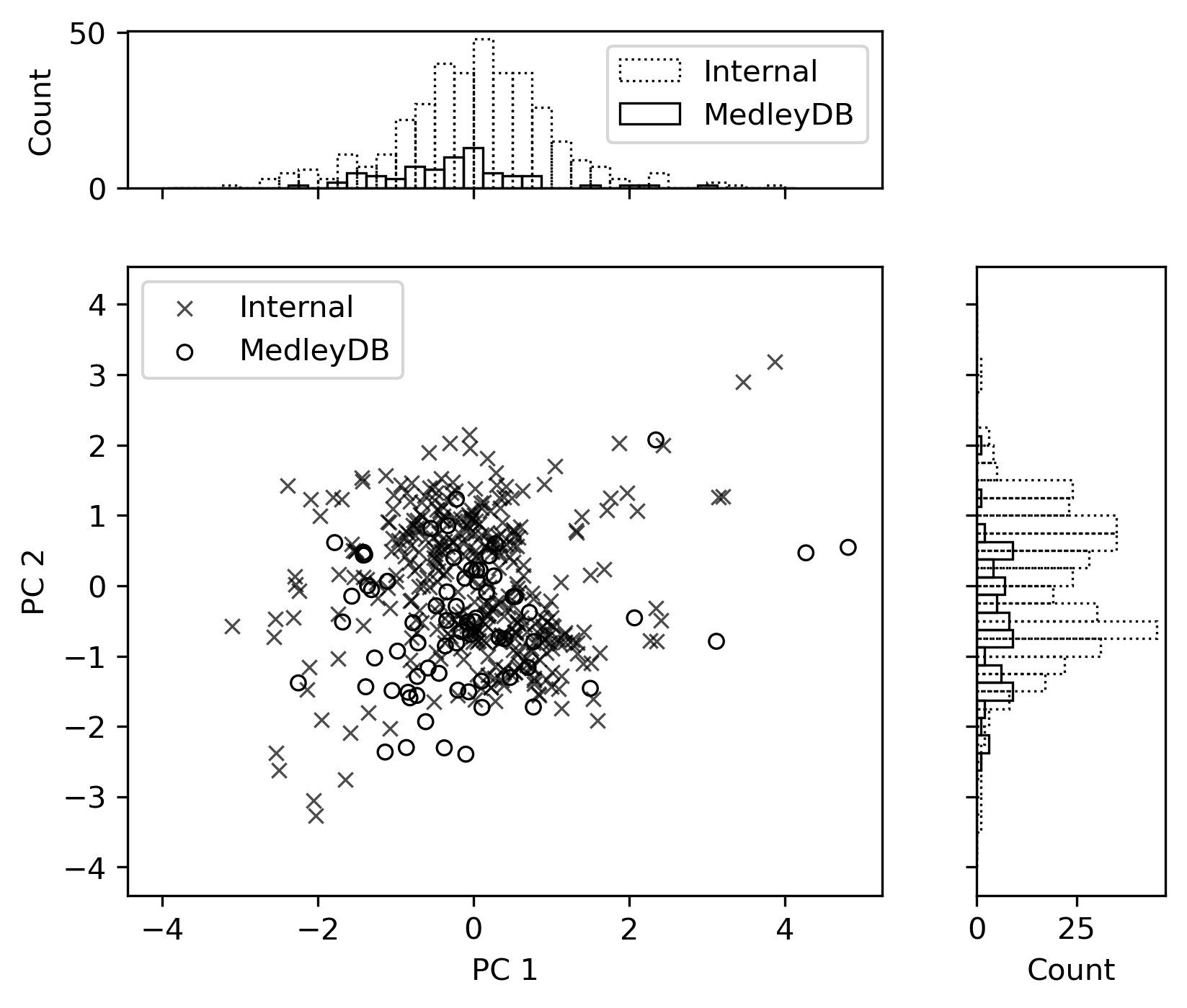}
    \caption{The scatter plot of the first two PCs' weights using the PCA model of \texttt{Internal}.}
    \label{fig:pca12}
\end{figure}

\begin{figure*}[tbh!]
    \centering
    \includegraphics[width=\textwidth]{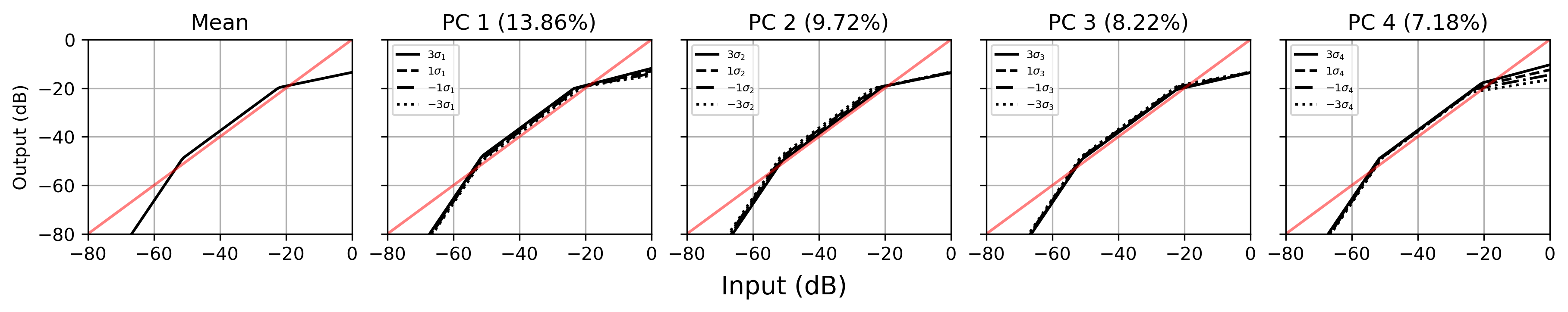}
    \caption{The compression curves (black) of the mean and the first four PCs of \texttt{Internal}'s PCA model.}
    \label{fig:comp_pc_compare}
\end{figure*}

\begin{figure*}[h]
    \centering
    \includegraphics[width=0.95\textwidth]{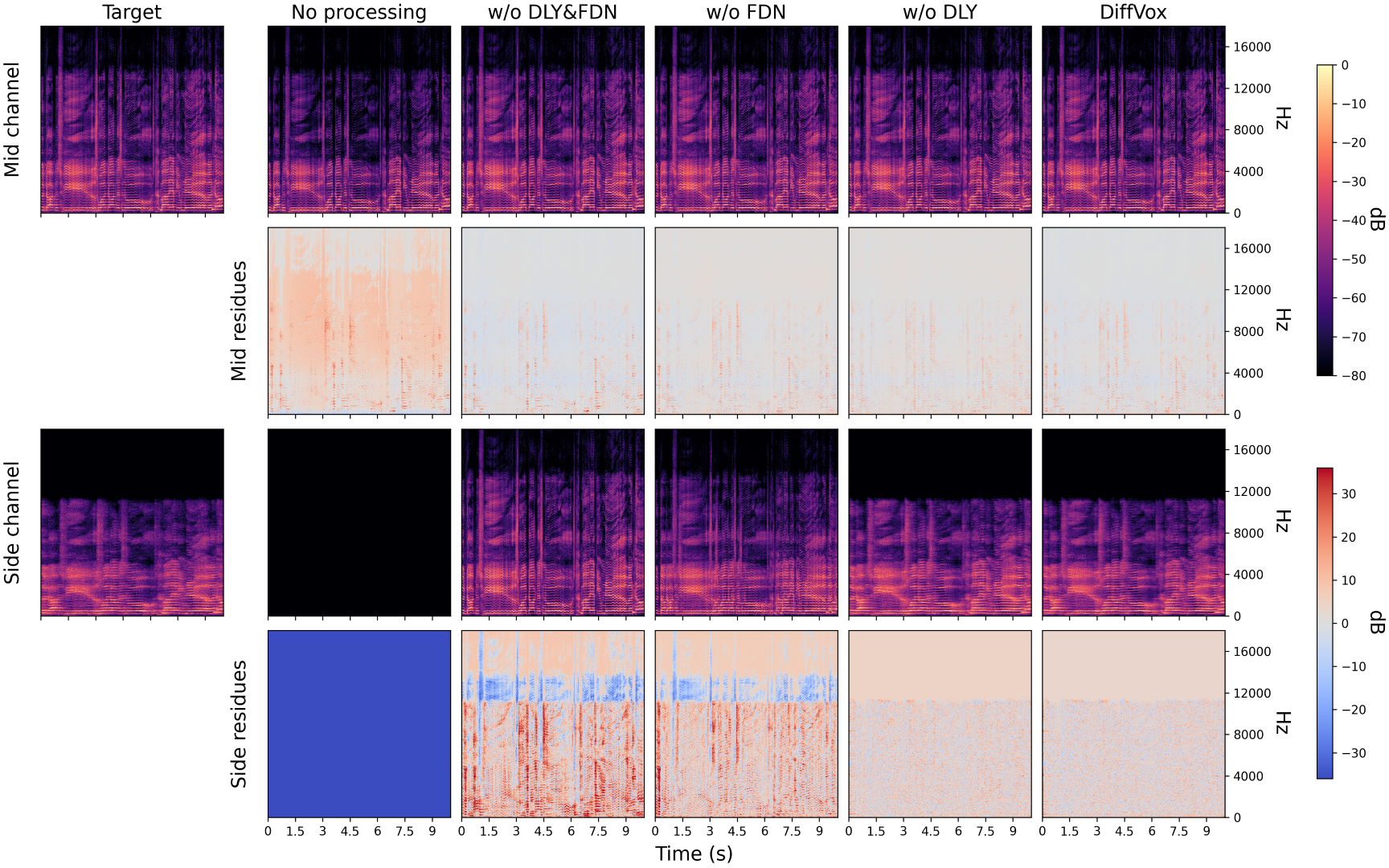}
    \caption{Matching the song \textit{Torres\_NewSkin} from \texttt{MedleyDB} using different effect configurations.}
    \label{fig:newskin}
\end{figure*}

\begin{figure*}[h]
    \centering
    \includegraphics[width=0.95\textwidth]{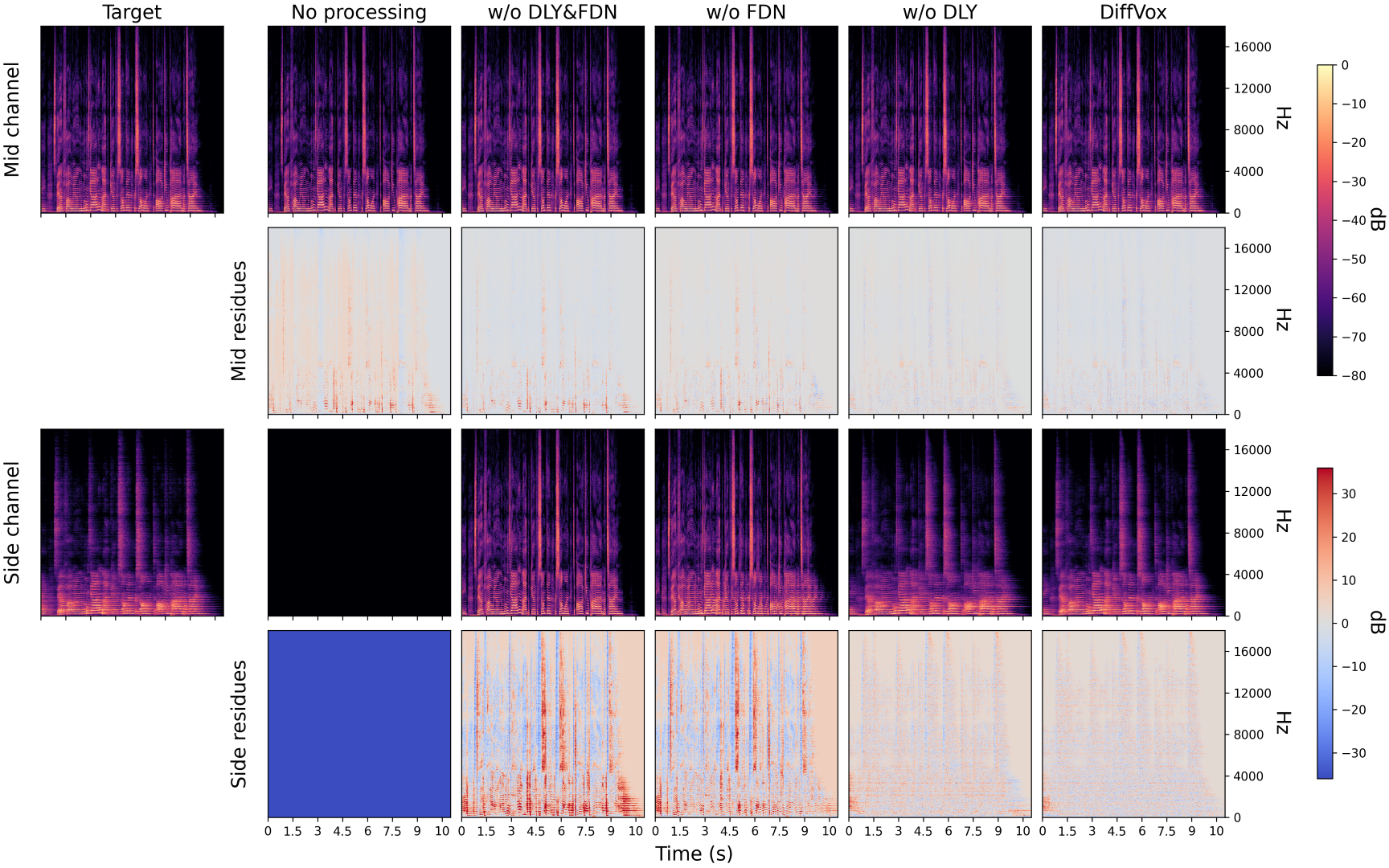}
    \caption{Matching the song \textit{MusicDelta\_Country1} from \texttt{MedleyDB} using different effect configurations.}
    \label{fig:country}
\end{figure*}

\begin{figure*}[h]
    \centering
    \includegraphics[width=0.95\textwidth]{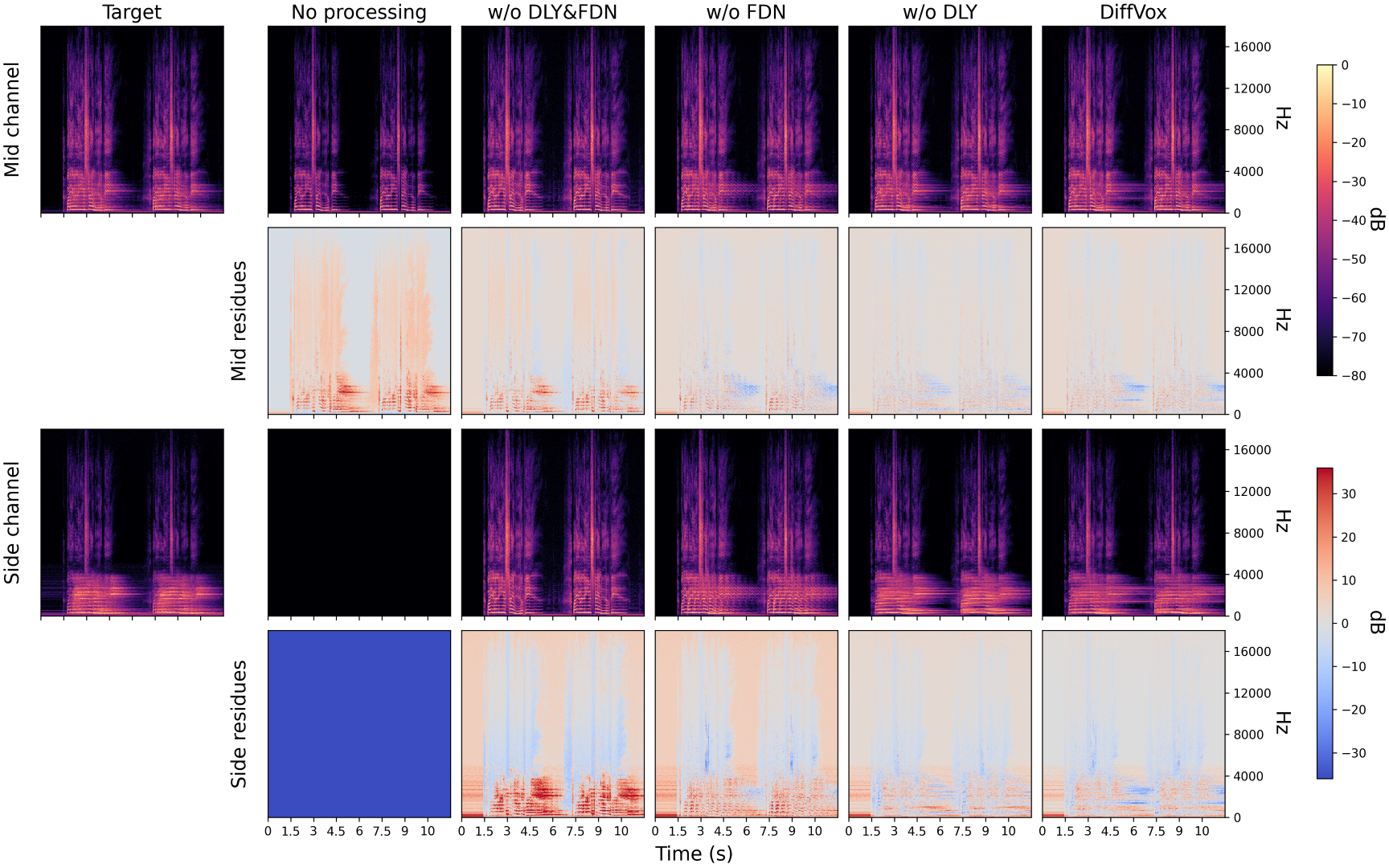}
    \caption{Matching the song \textit{StrandOfOaks\_Spacestation} from \texttt{MedleyDB} using different effect configurations.}
    \label{fig:spacestation}
\end{figure*}

\begin{itemize}
    \item Fig.~\ref{fig:panning_send} shows the histograms of the panning on the direct signal, $g_{\rm SEND}, \gamma_{\rm DLY}, g_{\rm DLY}$, and $d$ (delay time). For panning, 0 is the centre, -100 is hard left, and 100 is hard right.
    \item Fig.~\ref{fig:effect_corr} shows the correlation matrix of the individual effects in \texttt{Internal}'s effects model and the hierarchical clustering of the effects (Section~\ref{ssec:corr}). The hierarchical clustering is based on the correlation matrix with diagonal elements set to one. The colour of the cells indicates the strength of the correlation.
    \item Table~\ref{tab:internal_10_corr} shows the top ten most correlated parameters of \texttt{Internal} (Section~\ref{ssec:corr}). $f_{\rm LP}$ is the cut-off frequency of the low-pass filter, $g_{\rm PK2}$ is the gain of the second peak filter, DLY.LP means the low-pass filter in the delay effect, etc.
    \item Fig.~\ref{fig:pca12} shows a 2D projection on the first two PCs (divided by $\sqrt{\lambda}_{\boldsymbol{\theta}_i}$) of \texttt{Internal}'s PCA model.
    \item Fig.~\ref{fig:comp_pc_compare} shows the dynamic range compression curves of the compressor and expander.
    \item Fig.~\ref{fig:newskin}, ~\ref{fig:country}, and~\ref{fig:spacestation} show the spectrograms of the configurations stated in Section~\ref{ssec:match} on three songs from \texttt{MedleyDB} and their differences to the target sound.
\end{itemize}

\begin{table}[h]
    \caption{The top ten most correlated parameters of \texttt{Internal} and their correponding SCC from \texttt{MedleyDB}.}
    \centering
    \begin{tabular}{llr|r}
        \toprule
        \multirow{2}{*}{Parameter 1} & \multirow{2}{*}{Parameter 2}    & \multicolumn{2}{c}{SCC}                     \\
                                     &                                 & \texttt{Internal}       & \texttt{MedleyDB} \\
        \midrule
        $f_{\rm LP}$                 & $\gamma(e^{i\frac{44}{48}\pi})$ & 0.60                    & 0.32              \\
        $g_{\rm PK2}$                & $Q_{\rm PK2}$                   & -0.60                   & -0.10             \\
        $d$                          & $\gamma_{\rm DLY}$              & -0.58                   & -0.20             \\
        $f_{\rm LP}$                 & $\gamma(e^{i\frac{43}{48}\pi})$ & 0.56                    & 0.35              \\
        $CT$                         & make-up                         & -0.55                   & 0.06              \\
        $f_{\rm LP}$                 & $\gamma(e^{i\frac{45}{48}\pi})$ & 0.53                    & 0.19              \\
        $ET$                         & $ER$                            & -0.52                   & -0.30             \\
        $d$                          & $g_{\rm DLY}$                   & -0.51                   & -0.02             \\
        $\gamma_{\rm DLY}$           & $f_{\rm DLY.LP}$                & 0.49                    & 0.41              \\
        $g_{\rm FDN.PK2}$            & $f_{\rm FDN.PK2}$               & -0.46                   & -0.47             \\
        \bottomrule
    \end{tabular}
    \label{tab:internal_10_corr}
\end{table}

\end{document}